# Exploring Multifunctionality in MgO-Based Magnetic Tunnel Junctions with Coexisting Magnetoresistance and Memristive Properties

*Alejandro Schulman,* Elvira Paz, Tim Böhnert, Alex Steven Jenkins, and Ricardo Ferreira*

**Magnetic tunnel junctions (MTJs) and memristors are two key emerging nanotechnologies that attracted significant interest for potential applications at the forefront of the digital revolution, including sensing, data storage, and non-conventional computation. The co-integration of these phenomena into a single multifunctional device is an important step toward harnessing the re-programmability of memristive systems with the high yield and varied functionality of MTJs. This study demonstrates the co-existence of magnetoresistance and memristive properties on MgO-based MTJs. These devices show a magnetoresistance with a linear response as a function of a magnetic field and no hysteresis, which are the requirements for good magnetic field sensors, as well as demonstrating a non-volatile and quasi-analogue memristive behavior as a function of an applied electrical field down to nanosecond pulses. Furthermore, by doping the oxide barrier, the memristive power consumption is lowered by 20% giving the multi-functionality of the devices a promising scalability potential. This study also shows that, memristive switching can be reversibly used to completely suppress and recover the spintronic functionalities. These results can pave the way for a seamless co-integration of memristors and spintronic devices in complex reprogrammable circuits addressing applications such as reprogrammable multifunctional field sensor arrays and neuromorphic computing.**

## 1. Introduction

The interplay between magnetism and electrical conduction has led to the discovery of several interesting phenomena, such as giant magnetoresistance (GMR) in spin valves[1] and tunnel magnetoresistance (TMR) in magnetic tunnel junctions (MTJs).[2] These effects have revolutionized the field of spintronics which resulted in a wide range of applications such as magnetic sensors[3–5] and magnetic random access memories (MRAMs).[6,7] The ability to electrically control the resistance of a magnetic junction has led to the discovery of spin-transfer torque (STT), which allowed to lower the power consumption and increase their scalability potentiality.[8] This scalability potential and low power consumption, coupled with the non-linear magnetization dynamics, make MTJs increasingly attractive as building blocks in neuromorphic computing systems, particularly as neurons.[9] Furthermore, stochastic devices that utilize the intrinsic time randomness of STT reversal in a single magnetic tunnel junction have already demonstrated their potential for probabilistic computing applications.[10,11]

Another type of device that has gain attention lately is the so-called memristor, which exhibits resistive switching (RS) effect, which is characterized by a soft dielectric breakdown, which induces a reversible change of the resistance based on the application of electrical pulses.[12] Memristor devices are typically comprised of metal/insulator/metal junctions and, similar to the MTJs, are also regarded as one of the most promising devices for next-generation computing due to the possibility to obtain a quasi-analog control of the resistance and can be integrated in high density crossbar arrays,[13–15] making them very suitable to be utilized as synapses in neuromorphic systems.

These two classes of devices explore radically different physical mechanisms and their optimization process lead to the use of completely different classes of materials. The oxide in state-of-the-art memristors is made of a wide variety of materials such as $TiO_x$, $HfO_2$, $AlO_x$, $TaO_x$, and $WO_x$ with electrodes made of conductive materials such as Pt, Al, Cu, and W.[15,16] State-of-the-art MTJs on the other hand are make use of MgO insulating barriers encapsulated in CoFeB or FeB with very specific crystalline properties needed to ensure the coherent tunneling and high spin polarizations that result in high tunneling magnetoresistance ratios.[17] Despite the differences, both classes of devices are based on an insulating barrier encap-

A. Schulman, E. Paz, T. Böhnert, A. S. Jenkins, R. Ferreira
International Iberian Nanotechnology Laboratory, INL
Av. Mestre José Veiga s/n, Braga 4715-330, Portugal
E-mail: alejandro.schulman@inl.int

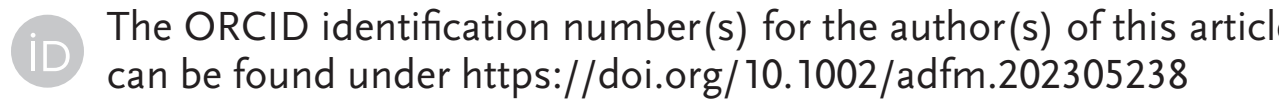
The ORCID identification number(s) for the author(s) of this article can be found under https://doi.org/10.1002/adfm.202305238

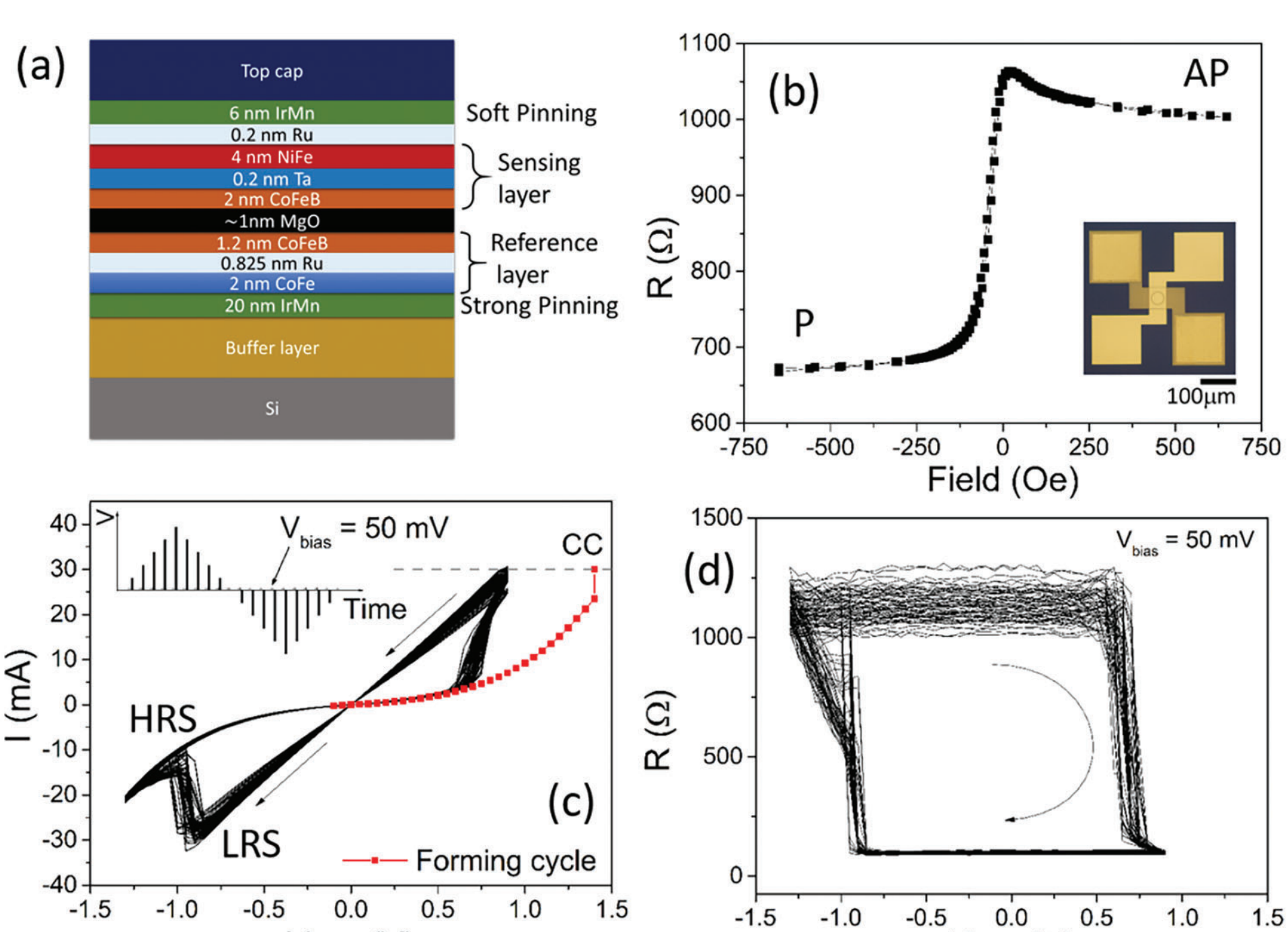

**Figure 1.** a) Schematic of the cross section of fabricated magnetic stack. The magnetic layers functionalities are labeled b) Magnetoresistance cycles for a MTJ with a diameter of 55 µm, this device has a TMR of 45%. inset: Top view optical image of a device. c) Pulsed Current–voltage (*I*–*V*) curve of the MTJ with the stack described in (a). The arrows indicate the direction of the voltage sweep. In red we show the first cycle, called electroforming process in which a higher positive bias needs to be applied. A schematic of the write-read pulsing scheme is depicted in the top left, pulse width was kept at 1 ms unless otherwise specified. The read pulses were performed 10 ms after the writing pulse with an amplitude of 50 mV. d) Resistance Switching Hysteresis Loop (RSHL) showing the value of the resistance measured in the read pulses vs the previous applied write voltage for 200 cycles.

sulated in conductive electrodes. So it is only natural to expect that memristive effects are present in magnetic tunnel junction.

In fact, recently there have been a lot of efforts focused at the electric field control of MTJs to tune their electronic[18] and magnetic properties,[19] aiming at the development of new multi-functional devices that exhibit multi-bit characteristics[20–28] and are equipped with the best qualities of both magnetic MTJs and memristors. However, previous studies that have investigated the RS effect in MTJs failed to demonstrate simultaneously both effects without the degradation of one of them. In most cases either the TMR is absent altogether or the resistance ratio of the memristive effect is greatly suppressed.[29–36] Here, we demonstrate both memristive and magnetoresistance effects in MTJ-based magnetic sensors without the degradation of either magnetic or memristive properties. Magnetoresistive sensors are of great interest for a varied range of technologies including automotive[37] and medical applications.[38] In our devices, we obtained a bipolar resistive switching ratio between the high resistance state (HRS) and low resistance state (LRS) up to two orders of magnitude, and the device shows a simultaneous TMR ratio of ≈45%, which is equivalent to the fabricated magnetic sensors without any memristive forming, promising a multi-functional devices with an electrically programmable magnetic sensor. Versatile reprogrammable MTJs can have an impact not just as magnetic sensors but can also potentially be of interest to other applications utilizing MTJs, such as MRAM and as building blocks of spintronic-based neuromorphic computing systems.[23,39–41] Furthermore, by doping the MgO barrier with Ta, we achieve 20% lower power consumption in the energy required to electroform the devices to exhibit memristive characteristics with the trade-off of dropping the TMR to 25%.

## 2. Results and Discussion

### 2.1. Magnetic Properties

We fabricated cylindrical magnetic tunnel junctions of different diameters ranging from 5 to 100 µm containing a soft pinned layer, this type of devices are typically utilized as ultra-sensitive magnetic field sensors.[42] A schematic of the stack configuration can be seen in **Figure 1**a. After fabrication, stacks were subjected to a 3-step annealing process.[43] The first annealing is performed at 330 °C in a 1T field for 2 h, this annealing induces the crystallization of the different layers and introduces uniaxial anisotropy in both the magnetic layers. The second annealing is necessary to rotate both pinning layers, but without rotating the crystalline magnetic anisotropy. So, a 1T field is applied for 2 h at 270 °C perpendicular to the first annealing. Finally, the third annealing step, parallel to the first annealing, rotates only the soft pinning layer with a 0.02T field, at 150 °C for 1 h.. As a result of our three-step annealing process, very good linearity with no coercivity is obtained, which is an essential characteristic of magnetic sensors. By changing the stack, the shape of the MTJ and the annealings we can tune the linear range of the sensor. The one presented in this work has a small linear range of ≈60 Oe.

We describe here in detail the results and characteristics of a device with a diameter of 55 µm, all studied areas show a similar resistance *x* area (RxA), TMR, and memristive properties indicating a uniform fabrication process. The influence of the area is discussed in section 2.2.1. A typical R(H) curve obtained under a small and constant current bias (0.1 mA) is presented in Figure 1b with a picture of a device. The transfer curve was obtained by applying an external magnetic field along the reference layer. Two clearly distinguishable magnetic saturation states exist at positive and negative applied magnetic fields, corresponding to antiparallel and parallel configurations, respectively. A tunnel magnetoresistance (TMR) ratio defined as $\Delta R/R_P = (R_{AP} - R_P)/R_P\ x\ 100$ where $R_P$ and $R_{AP}$ are the resistances when the magnetization of the reference and free layers are in the parallel and the antiparallel configurations, respectively. TMR of ≈45% was obtained at room temperature. When compared to similar devices routinely produced in the same lab,[42] these MTJs exhibit some differences with respect to the ideal behavior: i) the maximum resistance is not obtained for the maximum value of the external magnetic field and ii) the TMR is smaller than usual. The first deviation is attributed to changes made in the synthetic antiferromagnet (SAF) structure intended to increase the SAF thermal stability (important to keep the reference layer fixed when driving large currents through the nano-pillars which can easily heat the anti-ferromagnet up to the blocking temperature). The annealing protocol used to linearize the transfer curve was not optimized for this particular SAF and the reference layer is not perfectly collinear with the nominal direction along which the magnetic field is being applied, that is, when rotating the easy axis of the sensing layer to the orthogonal direction, the reference layer was slightly deviated from the its original orientation, and as result when large magnetic field are applied the sensing layer aligns with them in a direction that is not exactly the direction of the reference layer magnetization. The second deviation is attributed to the very large RxA of these devices. The Ion Milling conditions used to define the MTJ pillars must be tuned to remove any redeposited material in the pillar side wall that would shunt the current in a resistive channel parallel to the MgO barrier. In MTJs with RxA up to 10 kΩ µm$^2$ the conditions ion milling conditions used appear to be successful in removing this side-wall redeposited material. But in these devices with RxA≈2.4 MΩµm$^2$ it seems that not all the side-wall redeposited material is being removed and the resistive channel competes with the transport across the MgO barrier resulting in a reduced TMR. These changes with respect to the more conventional stacks were motivated by the need to produce devices where the memristive properties reported in this paper cannot be confused with other switching mechanisms also present in MTJ devices such as those associated with spin transfer torques or thermally assisted switching of the pinned layers. With additional optimization of the annealing and ion milling conditions targeted at this specific MTJ stack, it should be possible to produce devices with full TMR and perfect transfer curves. But these non-idealities have no impact in the results reported in this paper.

### 2.2. Memristive Properties

In order to find a memristive like behavior in these devices, they were subjected to a series of pulsed IV-sweeps in a sequence of steps 0 → $V_{max}$ → $V_{min}$ → 0, with a 10 ms low-voltage read between each step (schematized in the upper-left inset of Figure 1c). Current was measured both during writing and reading pulses. The during-pulse measurements represent the dynamic IV curve, while the read pulse measurements represents the remnant resistance state of the devices, which were also used when determining memristive switching ratios and voltage thresholds. We utilized this write and read approach as it gives us very useful information of the non-volatile effect of the electrical pulses as well as limiting the total heating and power that is delivered to the devices. The fabricated devices in its pristine state show a high resistance that depends on the area of the device with a $RxA_{AP}$ ≈ 2.4 MΩµm$^2$.

As with many types of memristor devices, it is necessary to perform an electroforming process at a relatively higher voltage (≈1.4 V) to create defects within the oxide barrier and jump-start the memristive behavior. This process was performed in the first cycle and is shown in red in Figure 1c. The device displays reversible, non-volatile bipolar resistive switching characteristics. After the electroforming process, the device switches from the as prepared state to the low resistance state (LRS), a compliance current (CC) is applied to avoid excessive current flow in order to avoid an irreversible breakdown of the devices. It is worth noting the LRS achieved by this method is much lower than the magnetic parallel configuration resistance state of the devices, indicating that we are not inducing a switch in the magnetization of the stack, similar to what is expected for STT devices. After the electroforming process, the voltage is swept between lower amplitudes (1 and −1.3 V) in a cyclic manner with the arrows in Figure 1c indicating the directions of the voltage sweeps.

When increasing the voltage, the current increases abruptly at ≈0.7 V which indicates the onset of the set process, which is how the transition toward the low resistance state (LRS) is commonly called. By reversing the polarity, the complementary behavior is observed at ≈0.85 V indicating the onset of the reset transition, which switches the device back to the HRS. When we repeated the IV sweeps within the same voltage polarity, we could not set and reset the device, confirming that our devices have a bipolar resistive switching. To further demonstrate the non-volatility of the effect, we applied 50 mV read pulses 10 ms after applying the write pulses. The results of the read cycles are plotted against the previous write pulse and is shown in Figure 1d. This type of plot is called resistance hysteresis switching loop (RHSL) and it gives more detailed information about the threshold voltages and remnant resistance values than the IV loops. In this case, the devices are very stable with a resistance change of 2000%, which is far larger than the resistance change of 45% due to the TMR effect defined in a similar way. To gain information about the cycle-to-cycle variation, the distribution of LRS and HRS is more intuitively displayed in **Figure 2**a. The Figure shows a clear switching window, with a higher dispersion on the HRS than the LRS, which is expected since the filament is not always 100% recovered in the reset process.[12] The inset of Figure 2a shows the distribution of $V_{set}$ and $V_{reset}$, devices exhibit a narrow distribution in both the $V_{set}$ and $V_{reset}$ with a standard deviation of 0.5 V in both processes, which is lower than the step utilized to measure the IV curve (0.1 V). It also can be noted the asymmetric values between $V_{set}$ and $V_{reset}$ with the $V_{reset}$ having slightly higher amplitude, a common signature of filamentary behavior. Additionally, a more comprehensive retention reliability test under 100 mV constant voltage stress were performed, results show that both states are stable, with a resistance ratio unchanged, showing no degradation during the studied time of about 24 h (Figure 2b).

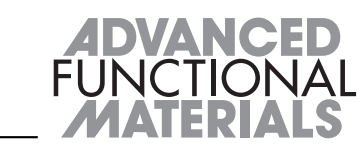

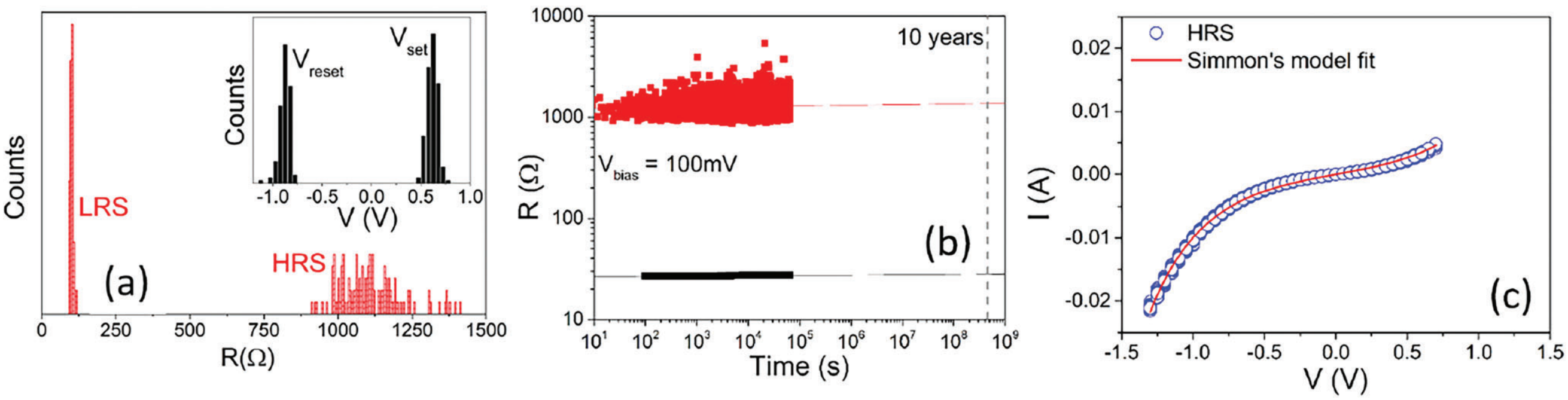

**Figure 2.** a) Distribution plot of the HRS, LRS showing cycle-to-cycle variations for 200 cycles. Inset: distribution of the $V_{set}$ and $V_{reset}$ voltages for the same cycles. b) Memory retention of HRS(LRS) red(black) for $\approx 10^4$ s at constant $V_{read} = 100$ mV. c) Fitting results for the HRS using the Simmons model.

To further understand the difference between both resistance states we look at the IV curves in more detail, from Figure 1c, it is clear that the LRS has a linear relationship between current and voltage, indicating an ohmic conduction mechanism, which is to be expected if the reason for the set process is a conduction path creation that short-circuits the tunnel barrier. On the other hand, the HRS show a typical nonlinear behavior associated with tunneling across a barrier. While several theoretical models have been proposed to describe tunneling transport of MTJs, the Simmon's model is the most widely utilized for MgO barriers showing symmetric IV characteristics.[44,45] This model gives a linear conduction regime at low voltages while for intermediate voltages (eV < $\phi$) the tunnel current is given by :

$$I = k_0 \frac{A}{t^2} \left[ \left(\phi - \frac{V}{2}\right) \exp\left(-k_1 t \sqrt{\phi - \frac{V}{2}}\right) - \left(\phi + \frac{V}{2}\right) \exp\left(-k_1 t \sqrt{\phi + \frac{V}{2}}\right) \right] \quad (1)$$

Where $k_0 = \frac{e^2}{2\pi h}$ and $k_1 = \frac{4\pi\sqrt{2me}}{h}$ Here, $A$ is the area of the tunnel barrier, $t$ is the tunnel barrier thickness, $e$ is the elementary electric charge, $m$ is the electron mass, $h$ is the Plank constant, and $\phi/2$ is the average between the barrier heights on both sides of the insulator barrier. Figure 2c shows the fitting results for HRS IV curves using Equation (1), where we obtained a value of $(1.3 \pm 0.2)$ nm for the barrier thickness ($t$) and $(1.07 \pm 0.01)$ eV for the barrier height ($\phi/2$), which are within the expected values for a high quality MgO barrier,[46] indicating that the original barrier is recovered in the reset process.

To further test the independence of the magnetoresistive and memristive effects, we performed a series of cycles under different magnetic fields, large enough to saturate the sensors. Results can be observed in **Figure 3**a where the RHSL are shown, the magnetic field has no observable effect over the memristive effect. We also measured TMR($H$) loops between the cycles both in the stopping in the HRS and in the LRS (marked with circles in the Figure 3a) to investigate how the resistive switching affects the magnetic properties of the stack. Three R($H$) cycles are presented in Figure 3b, a cycle measure before the electroforming of the device, a cycle after the set process in LRS and the last one after the reset process were the device is in the HRS and recovered its original resistance value. The R($H$) cycles show that in the LRS the TMR is totally suppressed indicating that the resistance of the resistive channel within the MgO is so low that the tunneling of electrons becomes negligible. Despite this, when the device is returned to the HRS after the reset process the TMR is recovered to its original value with no apparent degradation of the magnetic properties measured in the virgin state. This is a result that had never been demonstrated before. It is worth noting that the cycling of IV and corresponding R($H$) measurements has been performed for several cycles, showing systematic results without degradation, suggesting a robust multifunctionality of the devices. Having the possibility of switching a magnetic sensor on and off without degradation can be an attractive functionality for certain applications as magnetic sensors are usually utilized in bridges, having the possibility to short one can convert a magnetic sensor bridge into an magnetic field gradiometer.[47] But if this behavior can be extended to other classes of spintronic devices that are interconnected in complex circuits such as those used for neuromorphic applications,[48,49] it opens up a route for the implementation of reconfigurable spintronic circuits where individual devices can be suppressed upon fabrication effectively changing the topology and the functionality of the circuit after the fabrication process.

The RSHL curve in Figure 1d shows that the device can assume a range of resistance values between the HRS and the LRS values on both set and reset processes, demonstrating that the device is not only bipolar but also can be utilized in an almost analogue fashion. This is possible by carefully controlling the maximum applied voltage only in devices that don't possess abrupt transitions. We have achieved a quasi-analogue behavior in our devices (Figure 3c). To stabilize this intermediate states, we first set the device in the LRS with a 1 V pulse, we then performed half a pulsed IV ($0 \rightarrow V_{min} \rightarrow 0$) with decreasing $V_{min}$ in each cycle. $V_{min}$ was changed between −1 and −1.4 V in 0.01 V steps, giving a total of 40 curves. After performing this partial reset process, we measured a R(*H*) loop at low bias voltage to investigate the physical properties of these intermediate states. This analogue control of the resistance and magnetoresistance values will bring additional functionalities[50] when combined with other type of magnetic stacks, such as memory devices and spin torque nano-oscillators to create robust and behavior rich neuromorphic networks, in addition to the purely memristive functionalities already described in the literature such as matrix-vector multiplication[51] and convolution kernel operations.[52] Additionally, co-existence of TMR and memristive properties present in our devices could pave the way for a seamless co-integration of memristors and more complex spintronic devices in reprogrammable circuits where the connectivity of the devices can be changed modified after the fabrication process by an appropriate pulsing protocol, giving the devices a truly flexible characteristic from a circuit design standpoint. This flexible functionality could be of especially importance for neuromorphic computing applications where neurons and synapses need to be fabricated in close proximity and designing the same device that can behave as both is of great interest.

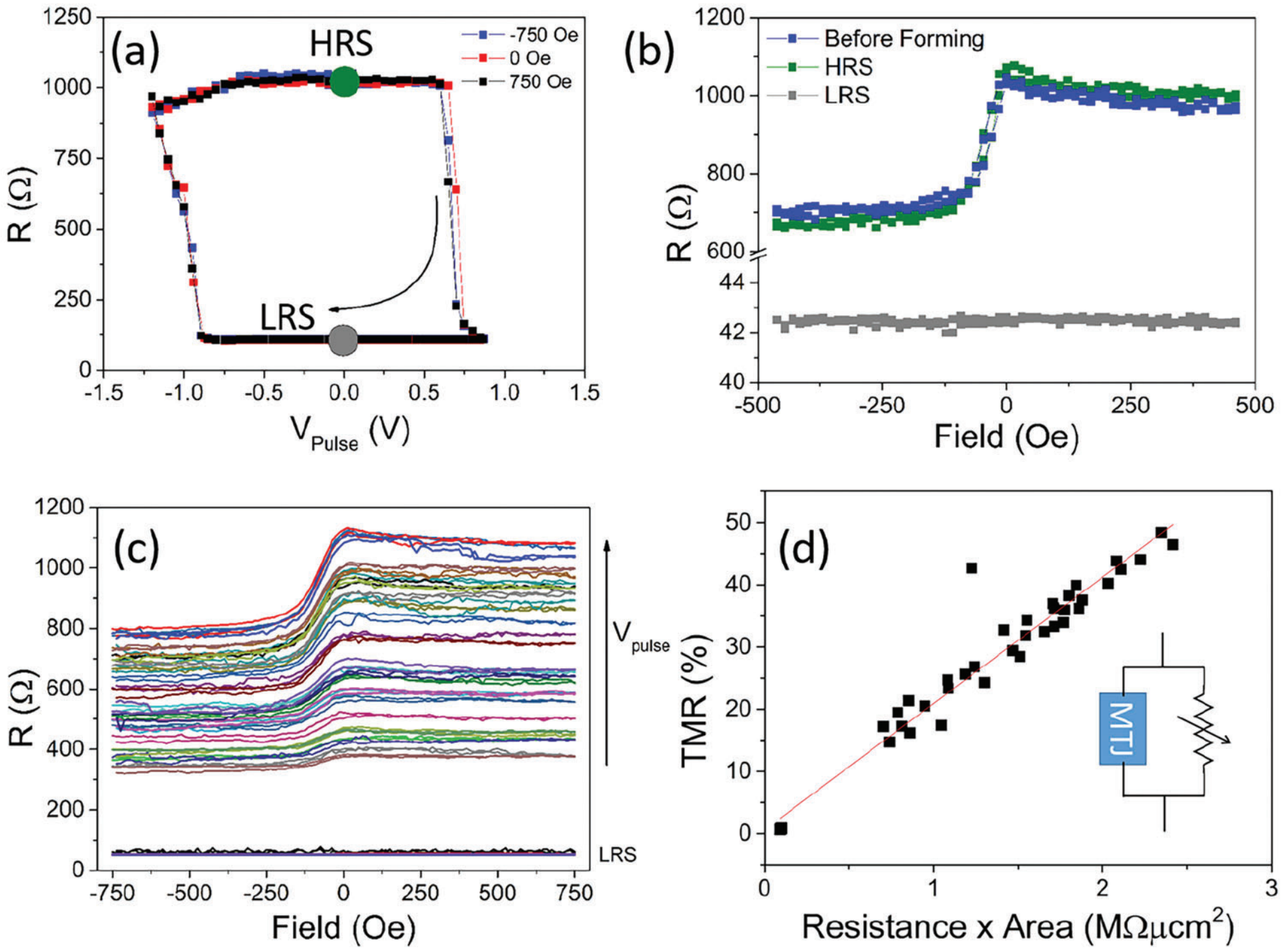

**Figure 3.** a) RSHL for a different applied external magnetic fields, the magnetic field has no effect on the memristive properties. The arrow indicates the direction of the sweep. The circles indicate the last step before performing the R(*H*) cycles in (b). b) Three R(*H*) cycles for the same device in different conditions: before the electroforming, in the HRS, and in the LRS. The LRS show no sign of magnetic interaction, indicating that the MgO is short-circuited, after switching back to the HRS, similar values to the original R(*H*) curve was obtained, proving that the barrier can be recovered in the reset process. c) R(*H*) curves for different intermediate states. The device was initialized in the LRS and then we applied a pulsed IV curve reaching different $V_{min}$ values, this operation performs a partial reset process, accessing a wide range of stable intermediate states. In each of these states we then perform the shown R(*H*) cycles. d) Calculated TMR of each of the intermediate states in function of the *RxA* calculated utilizing the measured parallel resistance. There is a clear linear relationship between this magnitudes, which is characteristic of having a resistance in parallel with the MTJ as schematized in the inset, the value of the parallel resistance is variable and depends on the $V_{min}$.

The lowering of the TMR effect in the intermediate states can be understood with a simple two parallel conduction channel model. In one channel, spin polarized electrons tunnel through the oxide barrier giving rise to the TMR effect and sensor-like magnetic field dependence while in the other conduction channel there is no spin polarization of the electrical current which gives rise to a pure resistance change without an impact on the magnetic properties of the tunneling channel.[53,54] The data in Figure 3d shows a linear correlation between TMR and the calculated resistance × area product which is the signature of a two-channel conduction mechanism where a spin-dependent tunneling path and a parallel spin-independent conduction path coexist. This is the reason the LRS shows no TMR signal, by creating a conducting path through the oxide effectively we shunted the barrier with an variable resistor that depends on the amplitude of the applied voltage bias, as schematized in the inset of Figure 3d.

#### 2.2.1. Area Dependence

Next, we investigate the effect of the junction area on the memristive properties. As shown in **Figure 4**a, the electroforming voltage has a slight area dependence, but more importantly, as devices get smaller (and more resistive) the total energy needed to perform the electroforming process is greatly reduced. Both $V_{Set}$ and $V_{Reset}$ have a slight size dependence, which is much smaller than the forming voltage. The remnant resistance measured at low bias for both HRS and LRS is presented in Figure 4b where it can be seen that the LRS is independent of the junction area, whereas the HRS is not and increases with decreasing the MTJs' area following the devices pristine resistances, which have a constant *RxA*. Those results indicate that the electrical conduction in the LRS is dominated by a metallic conduction path while the HRS is dependent on the whole oxide layer area.[30] This demonstrates that after electroforming, typically one single filament exists and that it can be switched by the bipolar pulses. In fact, given the results obtained by performing partial set and reset processes, it can be inferred that the size of the filament can be tuned in order to obtain a large range of resistance values in the junction. Given the characteristics of the conduction and the reversible characteristic of the effect, the most likely scenario is that the formed filament consists of conductive defects, more specifically oxygen vacancies, as it is a common mechanism for filament creation in oxide-based memristors. Furthermore, the presence of oxygen vacancies in MgO tunnel barriers is also well known.[55] For the larger devices, the resistance ratio is relatively small. This is due to the fact that the resistance of the pristine devices is comparable with the one of the created filament, since we can recover the original barrier area as shown by the IV curves, the resistance ratio significantly improves when the junction area is reduced (Figure 4c), since HRS is proportional to the area and LRS is not. In this respect, these devices show excellent scaling prospects. However, it is worth noting that due to the statistical distribution of these inhomogeneities where the filament is nucleated on the nanoscale, the properties of the devices may scale in a radically different manner in the nanometer range. Furthermore, moving toward nanometer size pillars implies reducing the MgO thickness to maintain the similar resistances values. In this new conditions, it is possible to switch the magnetization of the layer by current pulses by spin transfer torque effect. This new degree of freedom will need to be taken into account when designing the stacks in order to obtain devices that behave as we expected after the downscaling.

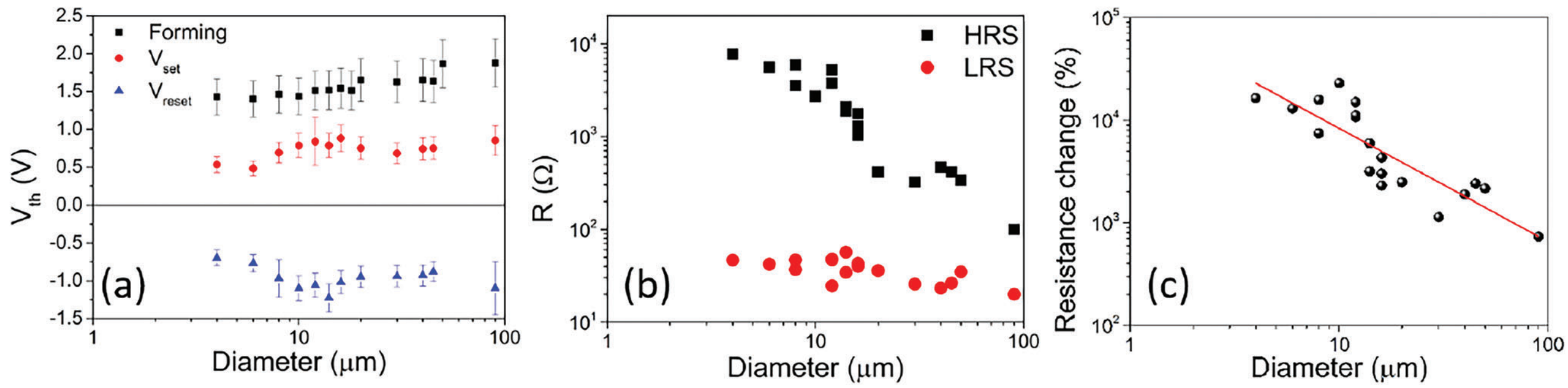

**Figure 4.** Area dependence of the memristive properties of the fabricated MTJs. a) Area dependence of the different threshold voltages (electroforming, set and reset). b) Area dependence of both the HRS and LRS, the LRS shows very weak area dependence due to the filamentary conduction. c) Resistance change (in %) in function of the MTJ diameter, as the LRS has a weak size correlation while the HRS scales with the area, the resistance change increases with the increase of the HRS.

### 2.3. Ta Doping for Energy Optimization

In addition to the previously discussed magnetic stack, we also fabricated a modified stack with an ultra-thin layer of Ta particles inserted in the middle of the MgO tunnel barrier. Doping of the oxide layer in memristors is used to tailor the properties of the devices and it has a very strong effect. Here we investigated the possibility of using a similar mechanism in MTJs using as a dopant one of the typical materials already used in the MTJ stack. Ta was chosen, but for no particular reason. It is possible that other types of materials are a better choice for this particular end.

To explore this, we inserted a Ta layer with a nominal thickness of 0.06 nm in the middle of the MgO barrier, resulting in a CoFeB/MgO/Ta(0.06 nm)/MgO/CoFeB structure. The addition of Ta in the barrier comes with a cost of magnetic performance, since we have now inserted defects into the MgO barrier. In **Figure 5**a, we can see how adding Ta lowers the TMR to 25% due to the inclusion of impurities and an increase of the *RxA* of the devices is also observed despite the thickness of the MgO layer remaining the same. Devices with and without the inclusion of the Ta doping behave very similarly as memristor devices, in Figure 5b we show a comparison for a device of 50 µm diameter but all areas showed similar results. Devices both with and without Ta doping exhibit almost identical memristive behavior with similar forming, $V_{Set}$ and $V_{Reset}$ voltages. The main difference we observed between both stacks is that devices with the Ta doping have a higher pristine and HRS resistances and thus, required less current to operate, lowering the energy consumption by ≈20% (Figure 5c). This reduction in energy consumption per operation could be of major importance in applications like neuromorphic computing where millions of devices are expected to function in parallel and that the system performance scale with its size.

The values obtained in Figure 5c were calculated for writing pulses widths of 1 ms and in the forming process with as it is more energy consuming than either the set and reset processes. However, devices can switch at much faster speeds. In fact, we performed a RSHL for a 55 µm diameter device with a write pulse width of 10 ns (Figure 5d). For this, we utilized the stack with the Ta interlayer, and as a result, we achieved energy consumptions of ≈(350 ± 50) pJ for the forming process and an energy per operation during the set and reset processes of 5 and 100 pJ, respectively. **Table 1** presents a comparison of the specific characteristics for devices that also show both magnetoresistance and memristive effects, for this we defined $R_{ratio}$ as HRS/LRS. We also include the state-of-the-art performance of both pure MTJ and memristors for comparison, it is also worth highlight that these are the best performances for each category and no single device shows all together. For most of the works, although the as prepared devices show good TMR values (usually ≈100%), once the electroforming is performed the device will stop displaying TMR and it cannot be recovered in the reset process.[30,34,46] In these cases, one can choose once after fabrication which devices will be utilized for their magnetic properties or their memristive ones, but they are not reconfigurable. On the other hand, some works show simultaneous TMR and memristive effects, similar to our work, although in this case, the memristive performance suffers as a result where the highest achieved $R_{ratio}$ was of about 3 with a concurrent 10% in TMR.[31] It is also worth mentioning, that in most of these works, measurements were performed in a ramp measurements for the IV so extracting switching speeds and energy consumption is not relevant.

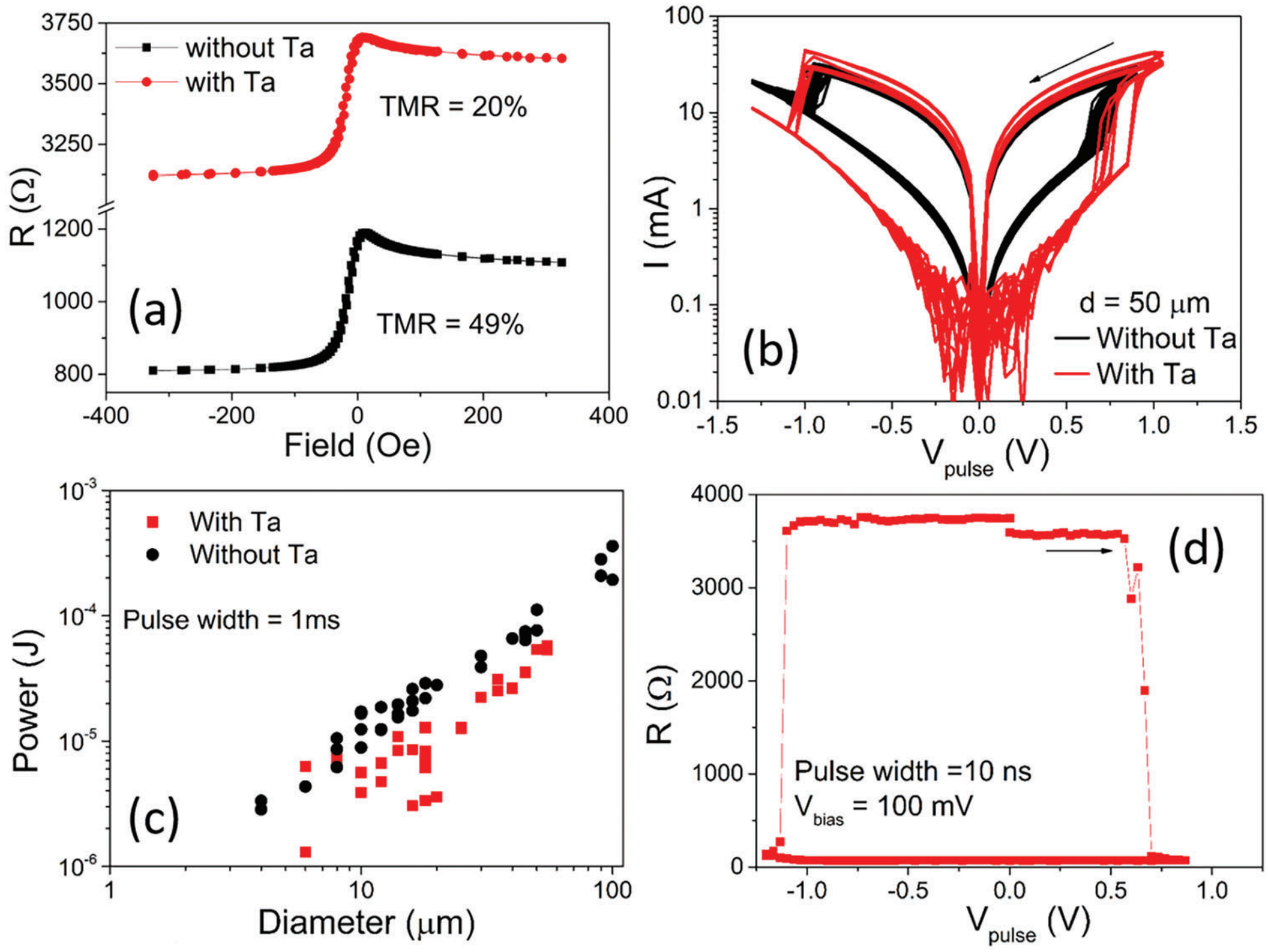

**Figure 5.** a) TMR (H) curves for MTJs of the same diameter (50 µm) for two different stacks, with and without the Ta impurities within the MgO barrier. b) *I–V* curves of the same stacks showing the memristive effect in the junctions. c) Energy consumption during the electroforming process for devices of different sizes both stacks. d) RSHL realized with a Ta embedded device with 10 ns writing pulses. The arrow indicates the sweep direction.

**Table 1.** Main characteristics of performance for devices that show both magnetoresistance and memristive effect. At the bottom we show state-of-the-art characteristics of pure MTJ and memristor devices.

| Materials | Multibit | TMR/$R_{ratio}$ | Operation voltage | Speed | Power consumption | Ref. |
|---|---|---|---|---|---|---|
| Fe/Cr/MgO/Fe | No | 3% / 3 | <1 V | NA | NA | [33] |
| CoFeB/MgO/CoFeB | Yes | No / 1000 | <2 V | NA | NA | [34] |
| CoFeB/MgO/CoFeB | No | 100% / 0.05 | <0.5 V | NA | NA | [32] |
| CoFeB/MgO/CoFeB | No | No / 3 | <2 V | NA | NA | [46] |
| Ti/Pt/MgO/Ta | No | No / 100 | <5 V | NA | NA | [18] |
| CoFeB/MgO/CoFeB | No | No / 3 | <2 V | NA | NA | [35] |
| NiFe/CoFe/$AlO_x$/CoFe | No | 10% / 3 | <1 V | NA | NA | [31] |
| MgO/$AlO_x$ | Yes | No / 30 | <1.5 V | 20 ns | ≈200 pJ | [30] |
| CoFeB/MgO/CoFeB | Yes | 45%/ 100 | <1.3 V | <10 ns | ≈100 pJ | This work |
| Memristor | Yes | 0% /$10^6$ | <0.5 V | <5 ns | <100 fJ | [56,57] |
| MTJ | No | 600%/ 1 | <1 V | <5 ns | <10 fJ | [56,57] |

The devices reported in this work show for the first time a quasi-analog control of the resistance of a magnetic tunnel junction using a memristor-like resistance switching of the MgO barrier which is non-volatile, fully reversible (even after the devices have been driven to a point where the TMR is fully suppressed), fast (down to 10 ns), with independent control of the memristive (resistance) and magnetic properties (coherent magnetization rotation in a sensor under an external magnetic field sweep) and very favorable scaling properties in the range of areas explored (down to 5 μm). This is a very encouraging starting point for researchers in the field to extend this work down to nm sized MTJs and other classes of MTJ-based devices like Spin Transfer Torque Nano-Oscillators with the potential of provide a mechanism to reconfigure spintronic devices and spintronic circuits upon fabrication using simply a mechanism that is already present in conventional MTJs with conventional materials.

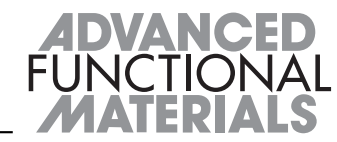

## 3. Conclusion

In summary, we have demonstrated that in MgO-based devices with a very conventional stack and materials, both memristive and magnetoresistive functionalities are observed without the degradation of either one. The room-temperature bipolar resistive switching ratio is over 10 000% while the TMR ratio is kept at ≈50% in HRS, which was the value observed in the devices before performing the electroforming process. In the LRS, devices show no magnetic response, suggesting the existence of two conduction channels with a spin-dependent tunneling path and a parallel spin-independent path. An almost continuous value of resistance could be obtained between a minimum resistance state and a maximum resistance which is similar to the pristine resistance value of the devices. This multilevel resistance value is consistent with a variable size conductive filament in parallel with the spin-dependent conduction path that typically exist on the MTJs as it has an effect of lowering the TMR proportionally to the resistance area. The area dependence of the HRS and LRS suggests that the switching mechanism is a filament-type resistive switching where the filament is most likely caused by a chain of oxygen vacancies within the MgO layer. By adding Ta doping to the MgO tunnel barrier, an important reduction of the power consumption of the memristive functionality was obtained with the drawback of lowering the TMR by adding a spin scattering center within the oxide barrier. The results lead to a promising possibility of integrating functionalities of MTJs with memristors and into the future spintronic multifunctional and reconfigurable devices and circuits.

## 4. Experimental Section

*Fabrication of the Devices*: Two magnetic tunnel junction stacks were deposited with the following stacks (thicknesses in nm):

(1) Buffer / 20 Ir20Mn80 / 2 Co70Fe30 / 0.8 Ru / 1.2 Co40Fe40B20 / MgO / 2 Co40Fe40B20 / 0.21 Ta / 4 Ni81Fe19 / 0.20 Ru / 6 Ir20Mn80 / capping layer
(2) Buffer / 20 Ir20Mn80 / 2 Co70Fe30 / 0.8 Ru / 1.2 Co40Fe40B20 / MgO / 0.06 Ta / MgO / 2 Co40Fe40B20 / 0.21 Ta / 4 Ni81Fe19 / 0.20 Ru / 6 Ir20Mn80 / capping layer.

They were deposited over 200 mm Si ⟨100⟩ wafers in a Timaris Singulus tool, the MgO thicknesses are not detailed because the deposition of the MgO in this particular tool is calibrated to obtain a target *RxA* instead of thickness. For the stack 1) a *RxA* of 2.4 M$\Omega\mu$cm$^2$ was targeted, while for stack (2) half of the MgO was deposited as in stack (1) then added the interlayer Ta before deposited the other half of the MgO oxide barrier. The MgO oxide was deposited from MgO targets without an oxidation step. Both stacks were then patterned into circular devices with diameters ranging from 5 to 100 μm. For this, a micro-fabrication method based on photo-lithography and ion milling of the MTJ pillars was utilized. Each pillar has four dedicated contact pads as can be seen in the inset of Figure 1b.

*Device Characterization*: All measurements were performed at room temperature. A small subset of as prepared devices (≈10% of each diameter) were characterized by measuring R(H) curves in a semi-automatic probe station equipped with a pair of Helmholtz coils that provide in-plane fields, a representative curve for this measurement is presented in Figure 1b.

For the memristive and combined measurements, the two-terminal pulsed IV characteristics were measured using a Keithley B2901B source-meter with the bottom electrode grounded in all the electrical measurements. For the TMR(*H*) loops a magnetic field in plane along the reference layer was applied by means of an electromagnetic coil with an U-shape iron core while the Keithley B2901B was utilized to measure the resistance of the devices at a low bias current. For the nanosecond pulses, an arbitrary waveform generator (Keysight M8190A) connected to an ZHL-4240-W+ amplifier was utilized while the Keithley B2901B was used to measure the resistance of the devices with a low bias voltage pulse 100 ms after the ns writing pulse.

## Acknowledgements

This work has received funding from the European Union's Horizon 2020 research and innovation programme under grant agreement No 101017098 (project RadioSpin), No 899559 (project SpinAge) and No 101070287 (project Swan-on-chip).

## Conflict of Interest

The authors declare no conflict of interest.

## Data Availability Statement

The data that support the findings of this study are available from the corresponding author upon reasonable request.

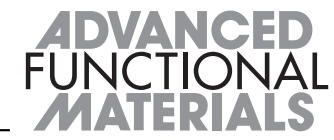